\documentstyle[psfig]{caosp}
\begin{document}
\pubyear{1993}
\volume{23}
\firstpage{7}
\hauthor{Glenn M. Wahlgren and Linus Dolk}
\title{Hot-Am stars as intermediary objects to the HgMn and Am stars}
\author{Glenn M. Wahlgren and Linus Dolk}
\institute{Atomic Spectroscopy Group\\Department of Physics, University of 
Lund\\S\"{o}lvegatan 14, S-223 62, Lund, Sweden}
\date{\today}
\maketitle
\begin{abstract}
We describe the early results of an investigation into the spectral 
characteristics of the hottest Am stars (A0 - A2) in an attempt to link 
the HgMn and Am classes of chemical peculiarity.  A limited sample of hot-Am 
stars was searched for the presence of lines from the very heavy elements 
platinum and mercury, as well as a search for lines from the rare-earth 
elements in HgMn stars. Our analysis of the strong optical platinum and 
mercury lines in the spectrum of the HgMn star HR 7775 has detected isotopic 
shifts that are different from those found at ultraviolet wavelengths and, in 
the case of mercury, vary with ionization stage. 
\end{abstract}
\section{Introduction}
The creation of stellar groupings by spectral characteristics is an often
necessary first step in being able to explore underlying physical
processes. The descriminating, or classification, criteria are often
chosen arbitrarily from available observational data and without regard for
the physical mechanisms which create them. Thus, a single physical process may 
be responsible for the variation of spectral characteristics
over a great range of temperature or luminosity.
In some sense, the warm chemically peculiar stars 
may be overclassified.  Various subclasses exist which are loosely arranged by 
temperature or enhanced element, but
they are now beginning to loose their identity and purpose as additional 
elements and spectral regions are explored.

We describe here the analysis procedure and early results of a project that 
has been undertaken to ascertain the extent of common spectral properties 
amongst warm chemically peculiar star groups. Our motivation is to test the 
theoretical hypothesis that diffusive element separation in the stellar 
atmosphere acts throughout the HgMn and Am star temperature range to create 
the abundance and isotopic anomalies observed (Michaud 1991). The concept 
that the Am and HgMn stars may be a related sequence  
can be traced back to Smith (1974) and perhaps even earlier, and has since
been reiterated by others. However, an insufficient amount of evidence
has been produced to either prove or disprove this assertion. 

The well accepted, but perhaps arbitrary, boundary at spectral type
B9.5/A0 separates the HgMn stars from the realm of the cooler, 
metallic-lined A (Am) stars. The classical Am stars
are usually limited to spectral types later than A3.
More recently the term {\it hot-Am} star has found its way into the literature,
representing those stars within the A0 - A3 subclasses whose spectra
display a certain degree of Am star spectral characteristics, most notably an
apparent deficiency of calcium and/or enhancements of iron-group element 
lines.  We are particularly interested in exploring the A0 - A2 stars for 
signs of an extension of the HgMn phenomenon to cooler temperatures as well as 
the B8 - A0 stars as an extension of the Am star phenomenon to hotter
temperatures.

\section{Defining intermediary objects}
If we are to recognize objects exhibiting both the HgMn and Am star phenomena
we must first consider the defining qualities of each of these classes.
Let us start by ignoring their classic definitions, as these were set up in an 
era that predates satellite ultraviolet observations, and is solely based upon 
the blue optical region. However, the first and second ions dominate the
spectrum for late-B through mid-A type stars, with most of their 
strongest transitions occurring at ultraviolet wavelengths. Therefore,
our thinking regarding the classification of peculiarity, and its implication
for the responsible physical processes, 
must be expanded to include the clues found at ultraviolet wavelengths.

For the cooler HgMn stars the defining characteristics must include the great 
line strength enhancements of the very heavy elements (VHE) Pt, Au, and Hg. 
Abundance anomalies for other elements, such as P, Ga, Mn, Tl, and Bi, have 
either been identified for a subgroup of stars or have yet to be thoroughly
investigated. From a variety of studies it can be
stated that the iron-group element abundances can be 
enhanced by up to 1 dex, which is also typical of the iron-group
elements among the Am stars. The identification
of rare-earth element (REE) lines in HgMn stars has, up to now, received
little attention. Guthrie (1985) noted 
the likely occurrence of lines from Nd II, Nd III, and Pr III
in the spectrum of HR 7775, but ``no other rare earths were found in HR 7775,
and no rare earth lines were detected in any of the other Hg-Mn stars''.

For the Am stars the relative weakness of calcium and scandium lines 
compared to those of iron, along with the presence of the REE 
are notable characteristics.  
Abundance enhancements for the heaviest elements (Pt, Au, Hg)
have not been identified in Am stars at optical wavelengths, which has lead 
to the general belief that these elements are not enhanced above solar-system
levels.

From these simplistic characterizations we have chosen to search for the 
presence of REE lines in HgMn stars and VHE lines in Am stars
as indicators of a common, albeit continuously varying, physical process 
that connects these two stellar groups.
Other common spectral properties are already known,
i.e. the abundance levels for light elements
(He, C, N, O) and iron-group elements, for example.
The literature provides few explicit examples of stars that display
spectrum features common to both groups. 

\section{Observational data}
Optical region spectral observations of HgMn, hot-Am, and Am stars were made 
at the 2.6-m Nordic Optical Telescope during 1996 November.
The SOFIN echelle spectrograph was operated at a resolving power of R = 
$\Delta$$\lambda$/$\lambda$ = 80000, with a single grating setting at blue
wavelengths including the lines Pt\,{\sc ii} 4046, 4061, 4288,
Au\,{\sc ii} 4052, Hg\,{\sc i} 4358, and Hg\,{\sc ii} 3984 \AA.
Each spectral observation consists of partial 
data from 14 contiguous orders. Typical signal-to-noise levels in the primary 
orders of interest were in excess of 100:1 and in some cases as large as 200:1.
Details of the observations and data reduction will be presented in a later
publication. We have supplemented the optical data with ultraviolet IUE
satellite high-dispersion spectra when available.

\section{The analysis procedure}

The detection and analysis of platinum and mercury lines in Am stars depends 
critically upon the use of synthetic spectra due to the crowded nature of the 
line spectrum in the blue spectral region. The Hg\,{\sc ii} $\lambda$3984
line is further complicated by being blended with the lines
Fe\,{\sc i} $\lambda$3983.7 and Cr\,{\sc i} $\lambda$3983.8 \AA.
The presence of an absorption feature at 3984 \AA\ can be traced from the
HgMn stars through the Am stars, but the relative contributions of the 
mercury, iron, and chromium lines varies with temperature
and abundance. Other mercury and platinum lines have their own blending 
concerns. For example, the strongest 
optical Pt\,{\sc ii} line, at $\lambda$4046 \AA, is blended with a weak
line of Hg\,{\sc i}.  Our line data for the platinum and mercury transitions 
accounts for their complicated hyperfine and isotopic structures, both from 
recent wavelength measurements with the Lund Fourier Transform spectrometer 
(Wahlgren et al. 1998a) 
and literature sources (Engleman 1989, Kalus et al. 1998).

We have determined the influence of blending lines by analysing the
spectra of 68 Tau and $\theta$ Leo. From IUE spectra of
Hg\,{\sc ii} $\lambda$1942 
one sees that these stars have only weak, if any, contribution from mercury.
Therefore, we do not expect to see the intrinsically weaker, optical 
region lines of mercury (or platinum). The rotational velocity 
and the abundances of Fe\,{\sc i} and Cr\,{\sc i} were determined from optical 
lines possessing experimentally determined oscillator strengths. These 
abundances were then used to evaluate the f-values of the blending lines to
the optical region platinum and mercury lines.

After establishing the atomic-line data we proceeded to synthesize the
spectra of the HgMn and Am stars. Model atmospheres were computed from the
code ATLAS9 based upon stellar atmospheric parameters derived 
from the uvby photometric calibration of Moon \& Dworetsky (1985).
The synthetic spectrum program SYNTHE and the atomic line data 
of Kurucz (1993) were used to compute the synthetic spectra, with the exception
of alterations that we have made for the heavy element line data, the
hyperfine structure of Y\,{\sc ii} 3982 \AA, and the $gf$-value for
Fe\,{\sc i} $\lambda$3983.7 \AA\ (O'Brien et al.1991).

\section{Early results}
The presence of rare-earth elements has previously been noted in the spectra 
of the cool, HgMn stars $\chi$ Lupi (Wahlgren et al. 1994) and HR 7775 
(Wahlgren et al.1998a) at optical wavelengths.  We therefore speculate 
that other, if not all, cool HgMn stars also display REE lines. Although it is
difficult to assess the exact abundance levels for the REE from
their third spectra, due to a lack of oscillator strength data, an 
abundance enhancement on the order of 1 to 2 dex is obtained from lines of
the singly-ionized state. This enhancement level is similar to that found
in Am stars.

Our high-resolution optical spectra have enabled us to study the isotopic 
shifts in the platinum and mercury lines found in the spectrum of the
HgMn star HR 7775. Observations of isotopic shifts represent important tests of
diffusion theory.  We were unable to satisfactorily fit the observed 
mercury lines with synthetic spectra that were computed assuming a mercury 
isotopic mixture based upon the q-formalism of White et al.(1976). 
By trial and error selection we were able to identify a difference in the
isotopic mixture between Hg\,{\sc ii} 3984 \AA\ and Hg\,{\sc i} 4358 \AA,
where the
former presents a mixture of $^{202}$Hg : $^{204}$Hg = 40:60 and the latter
line a terrestrial-like mixture. For each of the three Pt\,{\sc ii} lines an
essentially similar isotope mixture of $^{195}$Pt : $^{196}$Pt : $^{198}$Pt
= 10:60:30 provided good fits to the observed features. However, this 
platinum isotope mixture is different from that obtained from ultraviolet
transitions by Kalus et al.(1998). The potential significance of this 
difference lies in the
excitation energies of the transitions studied and the possibility that their 
isotopic variations (for Pt and Hg) are depth dependent in the stellar 
atmosphere. This would provide strong evidence for the effects of diffusion.
Further details are presented by Wahlgren et al.(1998a).

Our optical region search for VHE in hot-Am stars (Sirius, o Peg, HR 3383,
$\alpha$ Gem) has not yielded positive identifications.  The spectrum of 
HR 3383 did show promise for a mercury enhancement in the
Hg\,{\sc ii} $\lambda$3984 line at the 
detection limit. However, further analysis showed that the enhancement could 
be reproduced with an abundance variation of 0.05 - 0.10 dex for iron and 
chromium. The calculations show that for 
the hot-Am stars a mercury abundance enhancement of [Hg/H] = + 3 dex
would be easily noticeable in the 3984 \AA\ feature, with a minimum 
detection limit set at approximately [Hg/H] = + 2.5 dex.  As a further
check on the possibilty of a mercury enhancement in HR 3383, the IUE 
high-dispersion spectrum was investigated. The Hg\,{\sc ii} $\lambda$1942
line is present in HR 3383 at a strength
nearly identical to that found in Sirius ([Hg/H] = + 1.5, Wahlgren et al.
1998b). The three stars Sirius, HR 3383, and o Peg (Wahlgren et al. 1993)
all display the Hg\,{\sc ii} $\lambda$1942 line at a mercury enhancement 
of 1 - 1.5 dex. This value is remarkable for two reasons. First, if the
abundance enhancement level for the Am stars only approaches a level of
+1.0 to +1.5 dex, then by virtue of the difference in oscillator strength
of approximately 1.7 dex between the strong Hg\,{\sc ii} lines at 1942 and
3984 \AA, we would not expect to notice the presence of enhanced mercury at 
optical wavelengths. Thus, the B9.5/A0 boundary for the HgMn stars would be 
applicable as a classification criterion at {\it optical} wavelengths.
Secondly, the solar system abundances for platinum and mercury are 
poorly defined, the latter only from s-process systematics. 
This therefore raises the question of whether the abundance of platinum and 
mercury in hot-Am stars is a reflection of the galactic abundance at a
later epoch of star formation than that of the sun, or the result of a
reduced degree of diffusion relative to the HgMn stars.

\acknowledgements
GMW gratefully acknowledges a grant from the Royal Swedish
Academy of Sciences.


\end{document}